\documentclass[letter,twocolumn,showpacs,preprintnumbers,amsmath,amssymb]{revtex4}

\usepackage{graphicx}
\usepackage{dcolumn}
\usepackage{bm}
\begin{document}
\title{Antireflection Coating Using Planar Metamaterials }

\author{H.-T. Chen}
\author{J. Zhou, J. F. O'Hara, F. Chen, A. K. Azad, and A. J. Taylor}
\affiliation{MPA-CINT, MS K771, Los Alamos National Laboratory, Los Alamos, New Mexico 87545}
\date{\today}

\begin{abstract}
We present a novel antireflection approach utilizing planar metamaterials on dielectric surfaces. It consists of a split-ring resonator array and a metal mesh separated by a thin dielectric spacer. The coating dramatically reduces the reflectance and greatly enhances the transmittance over a wide range of incidence angles and a narrow bandwidth. Antireflection is achieved by tailoring the magnitude and phase shifts of waves reflected and transmitted at metamaterial boundaries, resulting in a destructive interference in reflection and constructive interference in transmission. The coating can be very thin and there is no requirement for the spacer dielectric constant. 
\end{abstract}

\pacs{42.79.Wc, 42.25.Hz, 78.67.Pt, 07.05.Tp}

\maketitle

When propagating light encounters an interface of two dielectric media with different refractive indices, some portion of energy is reflected back and the rest transmits through. This well-known phenomenon is the basis of numerous optical technologies and is described by Snell's law and Fresnel equations~\cite{Born_1980}. In many photonic applications however, reflection is undesirable and causes, for example, insertion losses or Fabry-P\'{e}rot fringes. Antireflection coatings using single or multiple layered dielectric films have long been the solution in the optical regime. Single layered quarter-wave antireflection coatings require a particular refractive index and a quarter wavelength thickness. Good performance is typically limited to a small range of incidence angles and a narrow bandwidth. This approach is scalable over a wide spectral range, but it is limited by natural material properties. For example, in the far infrared and microwave bands materials such as semiconductors and ferroelectrics often have large refractive indices, which makes it difficult to find appropriate coating materials that satisfy the index matching requirement. Furthermore, it is often quite challenging to fabricate high quality and relatively thick films required for long wavelengths. 

Antireflection is of additional importance when considering the generally low power in many far infrared or terahertz (THz) systems. There have been a few efforts to develop antireflection coatings at THz frequencies, although they are specific to certain substrates and/or coating materials. These approaches include: i) quarter-wave antireflection by optical lapping~\cite{Kawase_AO_2000}, various polymer coatings\cite{Englert_GRS_1999,Gatesman_AO_2000,Lau_AO_2006}, and vacuum or CVD deposition ~\cite{Fenner_MRSSP_2007,Hosako_AO_2003}; ii) fabricating artificial dielectric structures (surface relief structures) to achieve impedance matching with a gradient index~\cite{Chen_APL_2009,Bruckner_OE_2007} or by creating an effective index value satisfying quarter-wave antireflection~\cite{Wagner-Gentner_IPT_2006}; and iii) coating with very thin metal films having very precisely controlled thickness~\cite{Kroll_OE_2007,Thoman_PRB_2008}. All these approaches suffer from severe limitations; For instance, surface relief structures demand special fabrication procedures, in addition to the fact that substrate surfaces are no longer flat or smooth, limiting its application to many photonic devices. 

Antireflection coatings operate by overcoming the mismatch between two materialsÕ intrinsic impedances $Z_i = \sqrt{\mu_i / \epsilon_i}$ , where $\mu_i$ and $\epsilon_i$ are the media electric permittivity and magnetic permeability, respectively. Metamaterials with independently tailored $\epsilon$ and $\mu$~\cite{Smith_PRL_2000,Shelby_Science_2001} have inspired new potential applications including electromagnetic invisibility~\cite{Lee_OE_2005,Pendry_Science_2006} and enhanced light absorption~\cite{Landy_PRL_2008,Tao_OE_2008}. In principle, by tuning the effective $\epsilon$ and $\mu$ of metamaterials, it is possible to match impedances of two different media and thereby alleviate the non-availability of natural materials with the required index of refraction. But so far the general concept of metamaterials still suffers from high loss. Three-dimensional fabrication is another challenge. In this Letter, we present a novel metamaterial antireflection coating that dramatically reduces the reflectance from a dielectric interface and enhances the transmittance over a wide range of incidence angles at any specified wavelength from microwave to mid-infrared, and, as a proof of concept, our work has been focused on the THz frequency range. We have successfully modeled and identified the underlying mechanism through analytical derivations and finite-element numerical simulations.

\begin{figure}[t!]
\centerline{\includegraphics[width=2in]{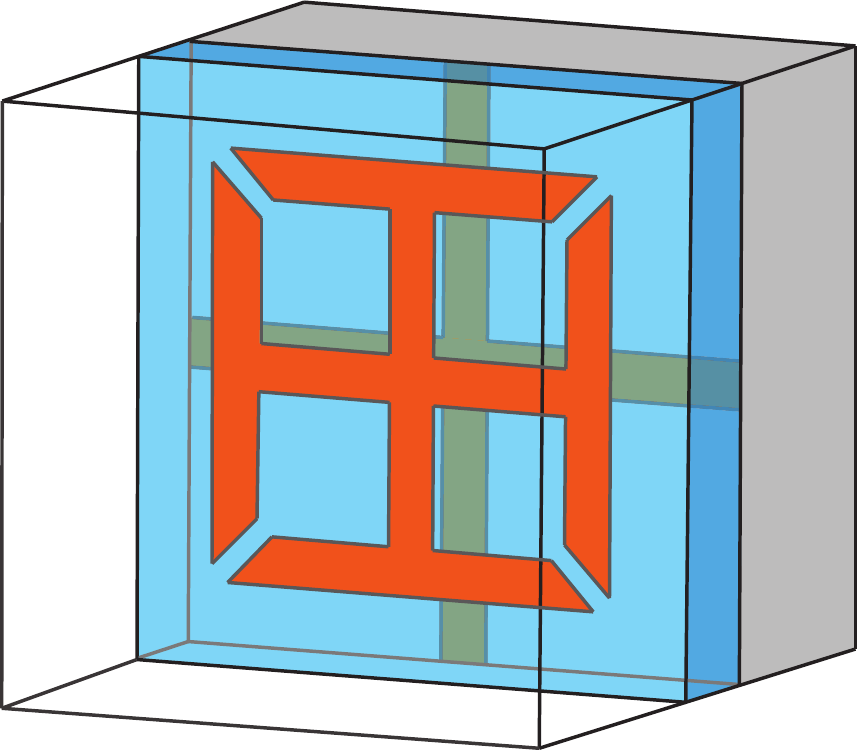}} \caption{ (color online). Schematic design of the metamaterial antireflection coating. From front to back: air, SRR, spacer, mesh, and substrate. This design is also used in the numerical simulations.} \label{Fig1}
\end{figure}

The THz metamaterial antireflection coating consists of an array of gold electric split-ring resonators (SRRs)~\cite{Chen_OE_2007} and a gold mesh~\cite{Ulrich_IP_1967} patterned using conventional photolithography methods, and separated by a spacer layer of spin-coated and thermally cured polyimide (dielectric constant $\sim$3.5)~\cite{Tao_OE_2008}. The four-fold symmetric design makes it polarization insensitive under normal incidence. The 1~cm $\times$ 1~cm coating was fabricated on an intrinsic gallium arsenide (GaAs) substrate, with a unit cell schematically shown in Fig.~\ref{Fig1}.  In this design, the width of the metal lines is 4~$\mu$m, and the SRRs have outer dimensions of 36~$\mu$m, 4~$\mu$m gaps, and 46~$\mu$m periodicity. The polyimide spacer thickness is expected to be about 13~$\mu$m. The metamaterial antireflection coating was characterized at various incidence angles through reflection and transmission measurements using a fiber-coupled THz time-domain spectrometer~\cite{OHara_JNO_2007}. Measurements were performed with both TM (transverse magnetic) and TE (transverse electric) incident waves.

\begin{figure}[b!]
\centerline{\includegraphics[width=3in]{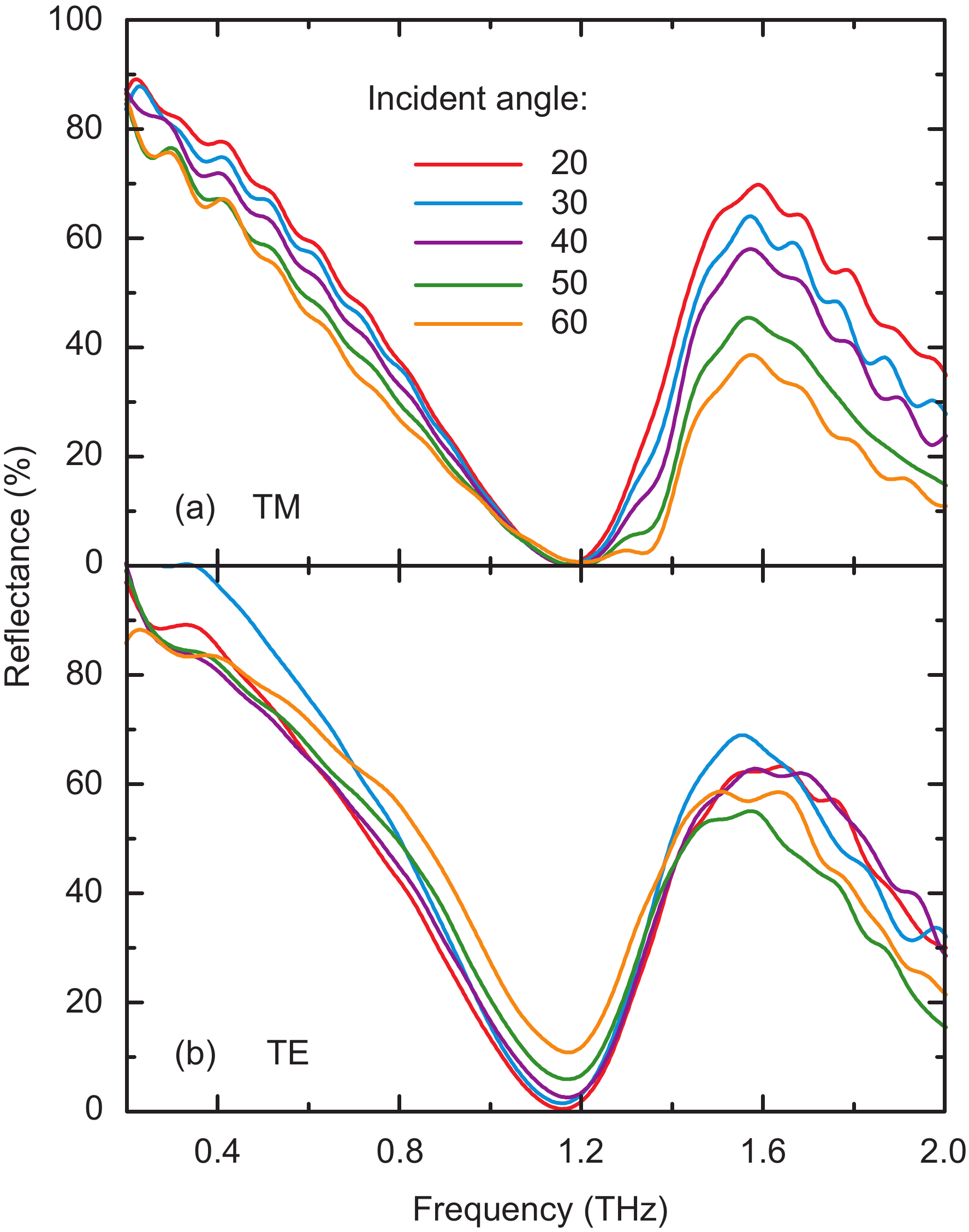}} \caption{Experimental results of reflectance from the metamaterial-coated GaAs surface for (a) TM and (b) TE polarized incident THz radiation, respectively, at various incidence angles.} \label{Fig2}
\end{figure}
The frequency dependent reflectance is shown in Fig.~\ref{Fig2}, with measured incidence angles ranging from 20$^{\circ}$ to 60$^{\circ}$, limited by the mechanical layout of the THz spectrometer and the sample and illumination spot sizes. Figure~\ref{Fig2}(a) shows the reflectance spectra under TM polarization. At incidence angles of 20$^{\circ}$ and 60$^{\circ}$,the reflectance minima are 0.25\% and 0.64\%, respectively. For a bare GaAs surface, the reflectance values are 27.2\% and 8.7\%, respectively. The lowest reflectance of 0.005\% is achieved at the incidence angle of 47.5$^{\circ}$. Figure~\ref{Fig2}(b) shows the reflectance under TE polarization at various incidence angles. The reflectance minimum is 0.46\% at 20$^{\circ}$, and it gradually increases to 10.9\% at 60$^{\circ}$. Although the reflectance for TE polarization is larger than for TM polarization, particularly at increasing angles, it significantly reduces the reflectance relative to the bare GaAs surface, which yields 34\% and 56\% at 20$^{\circ}$ and 60$^{\circ}$, respectively. By optimizing the spacer thickness, the TE reflectance is expected to be similar to the TM reflectance, a fact which has been verified through numerical simulations.

\begin{figure}[b!]
\centerline{\includegraphics[width=3in]{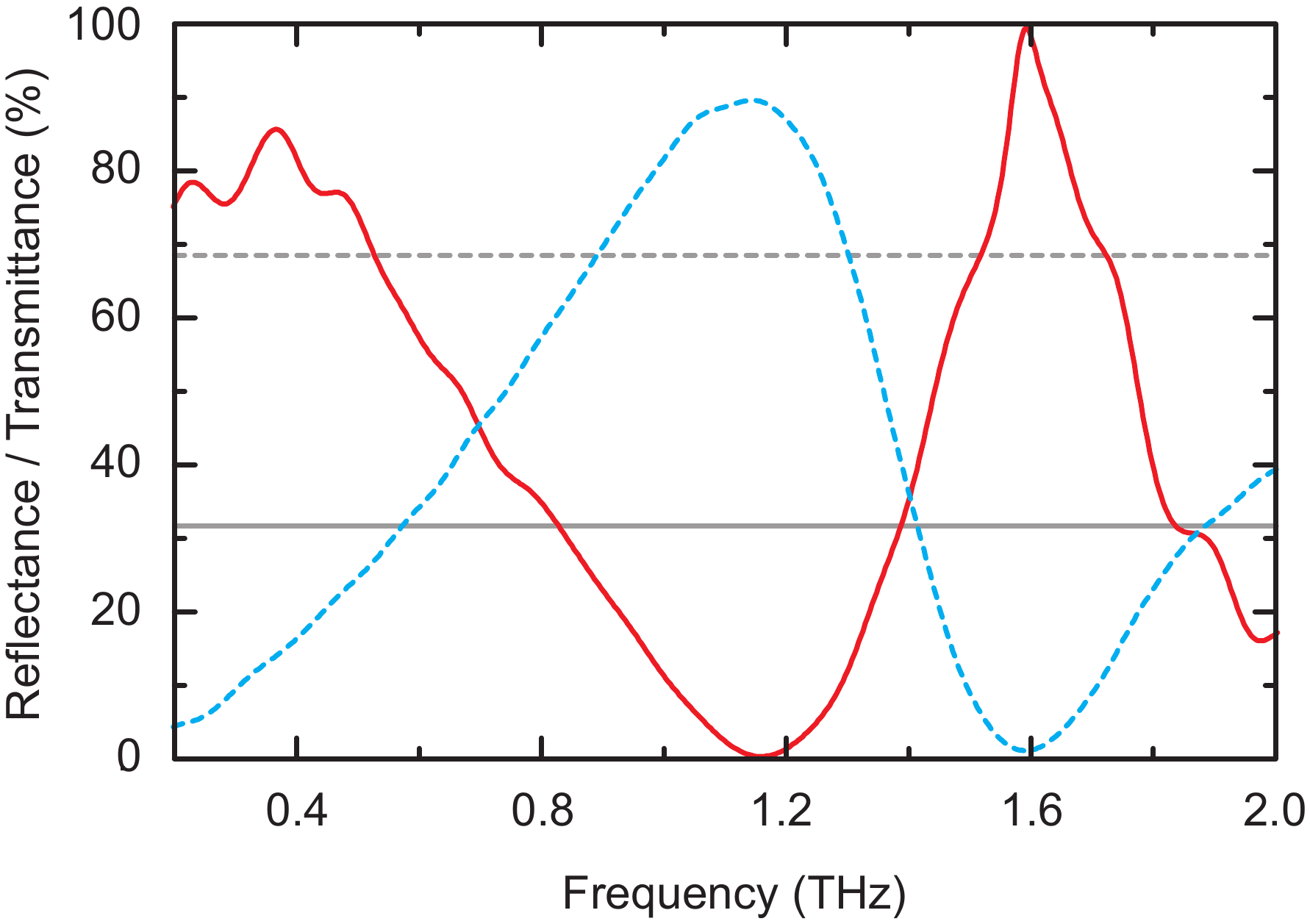}} \caption{(color online). Experimental reflectance (solid/red curve) and transmittance (dashed/blue curve) under normal incidence. The solid and dashed gray lines are the expected THz reflectance and transmittance, respectively, of a bare GaAs surface. } \label{Fig3}
\end{figure}
Since our THz system cannot directly measure the normal reflectance, it was determined by taking the advantage of echo pulses in the transmitted time-domain data. Echo pulses are generated by multiple reflections of the impulsive THz radiation between the two reflecting GaAs surfaces (with the front surface coated). As a result, the echoes measure the reflectance of THz radiation that is incident on the coating from the GaAs side, in contrast to the results in Fig.~\ref{Fig2}, where the incidence is from the air side. Figure~\ref{Fig3} shows a reflectance minimum of 0.32\% at 1.16 THz, consistent with the normal reflectance value expected  from the angular dependent data. This indicates that the metamaterial antireflection coating is effective for both light propagating directions, in contrast to, for example, thin metal film coatings where the antireflection can be achieved only when THz radiation propagates from the optically dense to less optically dense media~\cite{Kroll_OE_2007,Thoman_PRB_2008}. The normal transmittance was also measured, from which we obtained a transmittance maximum of 90\% as compared to 68\% through a bare air-GaAs interface. Without any optimization, this transmittance enhancement is already comparable with conventional quarter-wave coatings~\cite{Kawase_AO_2000,Englert_GRS_1999,Gatesman_AO_2000,Lau_AO_2006} or surface relief structures~\cite{Chen_APL_2009,Wagner-Gentner_IPT_2006} where 90\% or more transmittance was achieved. The loss of ~10\% is also significantly smaller than in very thin metal film coatings~\cite{Kroll_OE_2007,Thoman_PRB_2008} where the loss was $\sim$60\%. It is worth mentioning that our antireflection frequency is not far from the resonance frequency of the split-ring resonators resulting in appreciable loss, caused by both the metal and dielectric dissipation. Although we cannot substantially address metal losses, alternative lower loss dielectric spacer materials and redesigned non-resonant metamaterial structures should be able to further reduce loss and improve the transmittance. 

Finite-element numerical simulations were carried out using CST Microwave Studio 2009~\cite{CST2009}, to elucidate the roles of the SRRs, mesh and spacer in achieving good antireflection performance. The unit cell shown in Fig.~\ref{Fig1} was used in all simulations with a periodic boundary condition. Intrinsic GaAs was modelled as a lossless dielectric with $\epsilon = 12.7$ and gold was simulated as either a perfect electric conductor or with the default conductivity of $\sigma = 4 \times 10^7$~S~m$^{-1}$. Finally, the spacer layer was modelled as a dielectric material with varying thickness, dielectric constant, and loss tangent. Angular dependent numerical simulations are able to reproduce all experimental results shown in Figs.~\ref{Fig2} and \ref{Fig3}, which validates the modeling approach we will describe in the following. 

\begin{figure}[b!]
\centerline{\includegraphics[width=3in]{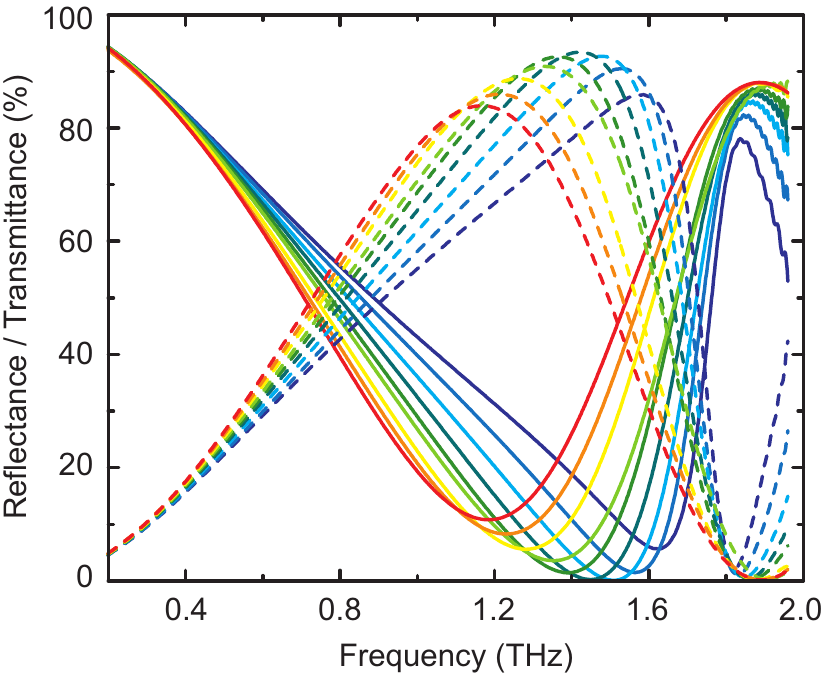}} \caption{Simulated reflectance (solid curves) and transmittance (dashed curves) under normal incidence as a function of spacer thickness varying from from 4~$\mu$m (dark blue) to 20~$\mu$m (red) with with an increment of 2~$\mu$m. Spacer dielectric constant is 1.5 and the loss tangent is 0.01. Default gold conductivity is used. } \label{Fig4}
\end{figure}
Figure~\ref{Fig4} shows the spacer thickness dependence of metamaterial antireflection. Although here the spacer dielectric constant of 1.5 was used, similar results are obtainable for any spacer dielectric constant, even ones larger than the substrate, as long as the loss is reasonably small. For any spacer index, there is always an optimized spacer thickness ($\sim$$\lambda/20$) where extremely small reflectance and enhanced transmittance can be achieved. It is worth noting that the spacer, by itself, could never function as a good conventional antireflection coating. Deviation from this optimized spacer thickness will result in a degraded antireflection performance and a shift in operation frequency. Upon further increase in the spacer thickness, the antireflection exhibits a periodic behavior (data not shown), which suggests a interference mechanism (destructive in reflection and constructive in transmission) rather than a magnetic response between the metal stripes of SRR and mesh~\cite{Zhang_PRL_2005,Shalaev_OL_2005}, which would produce a tuned $\mu$. This is further verified by the fact that antireflection performance is insensitive to the alignment between the SRRs and mesh. The simulations also reveal that losses in the metal and spacer material have a relatively small effect on the reflectance; instead such losses mainly degrade the transmittance.

\begin{figure}[b!]
\centerline{\includegraphics[width=3in]{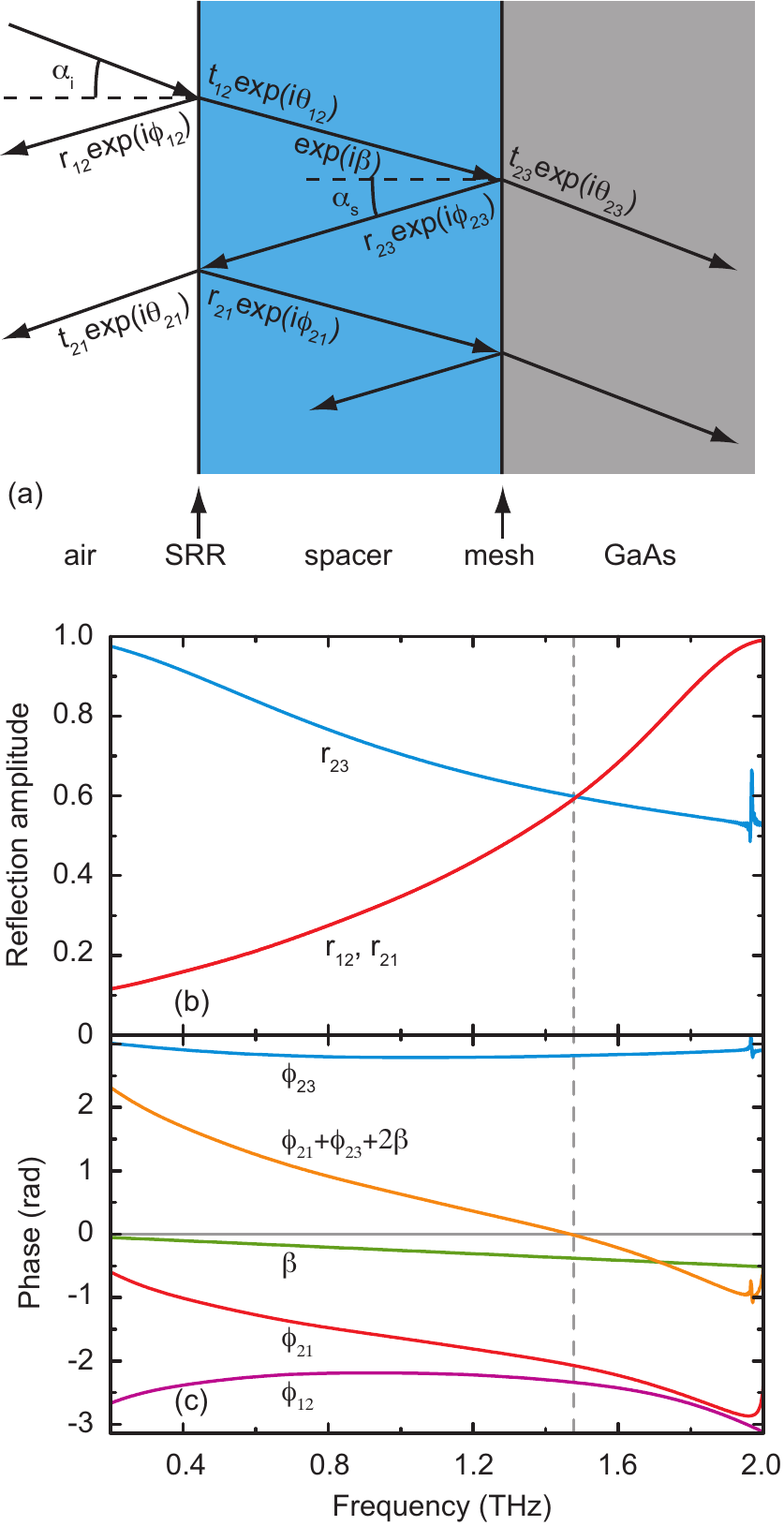}} \caption{(color online) (a) Illustration of interference model of the metamaterial  antireflection coating and associated variables.  (b) The magnitude and (c) phase of the complex reflection and transmission coefficients at the individual boundaries under normal incidence with the spacer dielectric constant 1.5 and without consideration of losses. The vertical dashed line indicates the optimized antireflection frequency. } \label{Fig5}
\end{figure} 
In order to further understand the antireflection mechanism, we model the SRR array and mesh as tuned-impedance interfaces with effectively zero thickness~\cite{Holloway_Metamaterials_2009,OHara_APEC_2007}. The modified air-spacer and spacer-substrate interfaces impart strong phase and magnitude shifts to the reflection and transmission coefficients. The whole antireflection system is depicted in Fig.~\ref{Fig5}(a), where the overall reflection and transmission are then superpositions of the multiple reflections and transmissions at the two interfaces. The overall reflection is given by~\cite{Born_1980}: 
\begin{equation}
\tilde{r} = \frac{r_{12}\exp{(i\phi_{12})} - r_{23}\exp{[i(\phi_{12}+\phi_{21}+\phi_{23}+2\beta)]}}{1 - r_{21} r_{23} \exp{[i(\phi_{21}+\phi_{23}+2\beta)]}},
\end{equation}
where all variables are illustrated in Fig.~\ref{Fig5}(a) indicating their meanings. Particularly, $\beta = - \sqrt{\epsilon_2} k_0 d / \cos{(\alpha_s)}$   is the phase delay as the wave propagates in one pass through the spacer, $\epsilon_2$ is the spacer dielectric constant, $k_0$ is the free space wavenumber, $d$ is the spacer thickness, and $\alpha_s = \arcsin{[\sin{\alpha_i/\sqrt{\epsilon_2}}]}$  for arbitrary incidence angle $\alpha_i$. The antireflection conditions are:
\begin{eqnarray}
r_{12} = r_{21} & = & r_{23},\\
\phi_{21}+\phi_{23}+2\beta & = & 2 m \pi,
\end{eqnarray}
where no loss has been assumed and $m$ can be any integer. 

The reflection and transmission coefficients at individual interfaces can be derived from numerical simulations of the isolated SRR and mesh layers, i.e., remove the irrelevant components in the unit cell. Under normal incidence, the magnitude and phase of the reflection coefficients are shown in Figs. 5(b) and 5(c). The overall spacer thickness dependent reflectance is then calculated from Eq. (1), and the transmittance is also similarly calculated. The results are in good agreement with those shown in Fig.~\ref{Fig4} if we take the losses into account. With the spacer thickness of about 10~$\mu$m, the antireflection conditions (2) and (3) are simultaneously satisfied at 1.48 THz, and we achieve nearly zero reflectance and 100\% transmittance. The mechanism is also responsible for the angular dependent reflection and transmission properties shown in Fig. 2, where antireflection conditions are satisfied at 47.5$^\circ$ resulting in the minimum reflectance and enhanced transmittance. 

In conclusion, we have experimentally demonstrated a novel metamaterial antireflection coating that dramatically reduces the reflectance and enhances the transmittance over a wide range of incidence angles with reasonable bandwidth, both for TM and TE polarization. The mechanism responsible for such antireflection has been identified. The SRRs and mesh modify the interfaces and impart strong phase and magnitude shifts to the reflection and transmission coefficients, resulting in the non-requirement for the coating index and very thin spacer thickness. Finally, the model indicates that any metamaterial structure, including many subwavelength frequency selective surface elements, either resonant or non-resonant, may be employed in this antireflection coating approach. This may further simplify the metamaterial structure and enable integration with previously demonstrated metamaterial devices~\cite{Chen_Nature_2006,Chen_Nature_2009,Padilla_PRL_2006,Chen_OL_2007,Chen_Nature_2008,Chan_APL_2009}, achieving novel functionalities for many THz applications.

We acknowledge support from the Los Alamos National Laboratory LDRD Program. This work was performed, in part, at the Center for Integrated Nanotechnologies, a US Department of Energy, Office of Basic Energy Sciences Nanoscale Science Research Center operated jointly by Los Alamos and Sandia National Laboratories. Los Alamos National Laboratory, an affirmative action/equal opportunity employer, is operated by Los Alamos National Security, LLC, for the National Nuclear Security Administration of the US Department of Energy under contract DE-AC52-06NA25396.
 
\bibliography{MyReference}

\end{document}